\documentclass[aps,amsmath,amssymb,amsfonts,twocolumn]{revtex4}
\usepackage{epsfig}
\usepackage{graphicx}
\usepackage{subfigure}
\usepackage{dcolumn}
\usepackage{bm}
\usepackage{amsthm}
\usepackage{enumitem}
\usepackage{slashed}
\usepackage{braket}
\usepackage{amsmath}
\usepackage{mathtools}

\makeatletter
\def\l@subsubsection#1#2{}
\makeatother

\newcommand{\bbC}{\mathbb{C}}

\newcommand{\calL}{\mathcal{L}}

\begin{document}

\title{Unified Lagrangians for gravity and gauge theory}

\author{Yasha Neiman}
\email{yashula@icloud.com}
\affiliation{Okinawa Institute of Science and Technology, 1919-1 Tancha, Onna-son, Okinawa 904-0495, Japan}

\date{\today}

\begin{abstract}
We present a new Lagrangian formulation of General Relativity with cosmological constant, coupled to Yang-Mills gauge theory. The formulation has a manifest color/kinematics-dual structure, both in the choice of fundamental fields and in the way they appear in the Lagrangian. The color/kinematics duality is between gauge and Lorentz generators. The fundamental fields consists of 1-form connections and 0-form curvatures -- a structure inspired by higher-spin gravity. The new formulation can be arrived at by combining appropriate versions of the Plebanski and Chalmers-Siegel Lagrangians, and integrating out the Plebanski 2-form. The construction comes in chiral and non-chiral versions. The chiral version is slightly simpler, and highlights the self-dual sector. The non-chiral one makes parity manifest, and doesn't force us to complexify the theory.
\end{abstract}

\maketitle

\section{Introduction} \label{sec:intro}

A longtime quest of theoretical physics is to maximize the similarities between Yang-Mills (YM) and General Relativity (GR), so that the former, easier theory may shed some light on the latter, more difficult one. In recent years, a large literature has appeared on ``color/kinematics duality'' in scattering amplitudes \cite{Bern:2008qj} and AdS/CFT correlators \cite{Armstrong:2020woi,Albayrak:2020fyp}. There, one finds that GR observables are often given by YM ones, appropriately ``squared'' through a procedure that analogizes YM gauge generators to the 4-momentum (the generator of diffeomorphisms). Such relations have also been studied at the level of spacetime fields \cite{Anastasiou:2014qba,Anastasiou:2018rdx}. In string theory, they are closely related to the KLT relations between open and closed strings \cite{Kawai:1985xq}. However, a full interacting action for GR+YM that treats both sectors on an equal footing (up to an appropriate ``squaring'' relation) has not yet been written. That is the goal of the present paper.

The simplest cartoon of color/kinematics duality is to consider the index structure of the metric $g_{\mu\nu}$ vs. the YM potential $A_{\mu a}$. Here, it's useful to think of spacetime indices $\mu,\nu,\dots$ as spanning the generators of translations (or diffeomorphisms), while the YM indices $a,b,\dots$ span the generators of gauge rotations. We immediately see that the metric $g_{\mu\nu}$ can be ``obtained'' from $A_{\mu a}$ by trading the gauge-generator index $a$ for a second translation-generator index $\nu$. Of course, what breaks this cartoon, and makes color/kinematics duality non-trivial, is that the roles of $g_{\mu\nu}$ and $A_{\mu a}$ in the Lagrangian are quite different: for instance, $A_{\mu a}$ is a connection, while $g_{\mu\nu}$ is a tensor.

At the same time, the Einstein-Yang-Mills Lagrangian contains a different hint of color-kinematics duality. Let us denote antisymmetric pairs of spacetime indices $[\mu_1\mu_2],[\nu_1\nu_2],\dots$ by 6d indices $I,J,\dots$, which we can think of as spanning the generators of \emph{Lorentz rotations}. Then the Riemann curvature and YM field strength can be written as $R_{IJ}$ and $F_{Ia}$, again exhibiting a color/kinematics pattern. Moreover, the Lagrangian now takes the form:
\begin{align}
    \calL = \frac{1}{\kappa}R^I{}_I - \frac{1}{2g^2}F^{Ia}F_{Ia} \ , \label{eq:L_cartoon}
\end{align}
where $\kappa = 8\pi G$ is the GR coupling, and $g$ is the YM coupling. The two terms in \eqref{eq:L_cartoon} clearly exhibit a color/kinematics structure, with the YM term involving an extra (unavoidable) ``squaring'' step. But, of course, this is still a cartoon: we ignored the special role played by the metric in converting between curved indices $\mu,\nu,\dots$ and flat Lorentz indices $I,J,\dots$, and in providing the volume form that must multiply \eqref{eq:L_cartoon}.

Our goal in this paper is to present a full-fledged formulation of GR+YM, with color/kinematics duality \emph{both} in the choice of fundamental fields \emph{and} in the structure of the Lagrangian (which will be built from expressions of the general form \eqref{eq:L_cartoon}). We were inspired to look for such a ``unification'' of GR and YM via the study of \emph{higher-spin gravity} -- the purported unified theory of massless fields of all spins and their interactions. In higher-spin theory, fields such as $g_{\mu\nu}$ and $A_{\mu a}$ are called \emph{Fronsdal potentials} \cite{Fronsdal:1978rb,Fronsdal:1978vb}. They work beautifully up to cubic vertices \cite{Sleight:2016dba}, but don't appear promising for defining the full theory. For that job, one uses ``unfolded'' formulations \cite{Vasiliev:1990en,Vasiliev:1995dn,Vasiliev:1999ba,Sharapov:2022faa,Sharapov:2022awp}, which consist of a 1-form gauge connection and a 0-form field strength. Here, ``unfolding'' refers to the fact that these 1-forms and 0-forms include not only the physical fields, but also towers of their derivatives. For some especially simple self-dual sectors \cite{Krasnov:2021nsq}, one can write standard Lagrangians without unfolding, but still with the basic ``1-form + 0-form'' structure. We thus set out to look for a similar formulation of GR+YM.

After the fact, our formulation follows from a simple (but overlooked) combination of ideas by Capovilla, Jacobson and Dell \cite{Capovilla:1989ac,Capovilla:1990qi,Capovilla:1991qb,Capovilla:1991kx}, see also Krasnov \cite{Krasnov:2011pp,Krasnov:2011up,Krasnov:2016emc,Krasnov:2017dww}. The idea is to combine the Plebanski Lagrangian for GR and the ``Chalmers-Siegel'' Lagrangian for YM \cite{Jacobson:1987yw,Chalmers:1997sg,Chalmers:1996rq} as in \cite{Capovilla:1991qb}, and then integrate out the Plebanski 2-form as in \cite{Capovilla:1991kx}. The simplest version of the construction is chiral, which is useful for treating the full theory as a perturbation over its self-dual sector. However, in Lorentzian signature, this requires us to use complex fields, making both parity and unitarity non-manifest. Fortunately, a real, non-chiral formulation is also possible. Both versions will make use of the special properties of the Hodge dual in 4d, so the construction is specific to 4 spacetime dimensions. We will present the non-chiral version first, followed by the chiral one, followed by an account of their origin from Plebanski + Chalmers-Siegel.

\section{Real, non-chiral Lagrangian} \label{sec:non_chiral}

\subsection{Index notations}

We work in a frame-like formalism, where all coordinate indices $\mu,\nu,\dots$ are subsumed into differential forms; in particular, our Lagrangians will be 4-forms. As in the vielbein formulation, we imagine a local tangent space at each point of the manifold with indices $A,B,\dots$, subject to the flat Minkowski metric $\eta_{AB}$. We will always use these indices in \emph{antisymmetric pairs}, for which we introduce composite \emph{bivector} indices $I,J,K,\ldots\equiv[A_1A_2],[B_1B_2],[C_1C_2],\ldots$, spanning the 6d space of antisymmetric $4\times 4$ matrices. This space is equipped with three structures: 
\begin{enumerate}
  \item A metric $\eta_{IJ}$ with $(3,3)$ signature (corresponding to spacelike and timelike bivectors). We'll use this to raise and lower indices.
  \item A Hodge-dual, parity-odd metric $\tilde \eta_{IJ}$, also with $(3,3)$ signature.
  \item The structure constants $f^I{}_{JK}$ of the Lorentz group.
\end{enumerate}
In vector indices, these structures are simply $\frac{1}{2}\eta_{A_1[B_1}\eta_{B_2]A_2}$, $\frac{1}{4}\epsilon_{A_1A_2B_1B_2}$, and $4\delta^{[A_1}_{[B_1}\eta_{B_2][C_1}\delta^{A_2]}_{C_2]}$. We denote the Hodge dual of any tensor (with respect to its 1st bivector index) as $\tilde T^{I\dots} \equiv \tilde\eta^I{}_J T^{J\dots}$. This satisfies $T_I\tilde U^I = \tilde T_I U^I$ and $\tilde{\tilde T}^I = -T^I$. 

Finally, we use indices $a,b,\dots$ for the YM gauge algebra, with structure constants $f^a{}_{bc}$. These indices are trivially raised/lowered using the Killing metric $\delta_{ab}$.

\subsection{Fields and Lagrangian}

Our fundamental fields are divided into 1-forms and 0-forms. The 1-form fields are the YM connection $A^a$ and the spin-connection $\omega^I$. From these, we construct curvature 2-forms as usual:
\begin{align}
    F^a &\equiv dA^a + \frac{1}{2}f^a{}_{bc}A^b\wedge A^c \ ; \label{eq:F} \\ 
    R^I &\equiv d\omega^I + \frac{1}{2}f^I{}_{JK}\omega^J\wedge\omega^K \ . \label{eq:R}
\end{align}
In addition, we introduce the 0-form fields $\phi^{Ia}$ and $\Psi^{IJ}$, which will be identified (through the equations of motion) with the YM curvature $F^a$ and the Weyl-curvature piece of $R^I$, respectively. The index symmetries we assume for $\Psi^{IJ}$ are:
\begin{align}
   \Psi^I{}_I = 0 \ ; \quad \tilde\Psi^{IJ} = \tilde\Psi^{JI} \ . \label{eq:Psi_symmetries}
\end{align}
These are some, but not all, of the symmetries of a Weyl tensor (in fact, they imply that the \emph{Hodge dual} $\tilde\Psi^{IJ}$ has precisely the symmetries of a Riemann tensor). The extra, non-Weyl-like components of $\Psi^{IJ}$ (i.e. the Ricci-like components of $\tilde\Psi^{IJ}$) will vanish under the equations of motion, but will contribute usefully to the variational principle. 

Our Lagrangian takes the form:
\begin{align}
    \calL[\omega^I,\Psi^{IJ},A^a,\phi^{Ia}] = -\frac{1}{2}M^{-1}_{IJ}\,V^I\wedge V^J \ , \label{eq:L}
\end{align}
where $V^I$ is a 6d ``vector'' of 2-forms, and $M_{IJ}^{-1}$ is the inverse of a 0-form matrix $M^{IJ}$. Specifically, $M^{IJ}$ and $V^I$ read:
\begin{align}
    M^{IJ}[\Psi^{IJ},\phi^{Ia}] &= \frac{1}{\kappa}\left(\frac{\Lambda}{3}\,\tilde\eta^{IJ} + \tilde\Psi^{IJ}\right) - \frac{1}{g^2}\,\tilde\phi_a^{(I}\phi_a^{J)} \ ; \label{eq:M} \\ 
    V^I[\omega^I,A^a,\phi^{Ia}] &= \frac{1}{\kappa}R^I - \frac{1}{g^2}\,\phi_a^I F^a \ , \label{eq:V}
\end{align}
where $\kappa = 8\pi G$ is the gravitational coupling, $g$ is the YM coupling, and $\Lambda$ is the cosmological constant. Note that both of the structures \eqref{eq:M}-\eqref{eq:V} follow a color/kinematics pattern similar to that of \eqref{eq:L_cartoon}. As we will see, the metric in this formalism will be contained in the 2-form $M^{-1}_{IJ}V^J$. 

\subsection{Equations of motion} \label{sec:non_chiral:equations}

Let us now analyze the equations of motion for the Lagrangian \eqref{eq:L}-\eqref{eq:V}, to see that it correctly describes GR+YM. The equations organize neatly in terms of the 2-form $M^{-1}_{IJ}V^J \equiv -\tilde\Sigma_I$ (the sign and Hodge dual here are for later convenience). Varying with respect to all the fields, we get the equations of motion:
\begin{align}
    \frac{\delta}{\delta\Psi^{IJ}} \ : \quad &\Sigma^I\wedge\Sigma^J \sim \tilde\eta^{IJ} \ ; \label{eq:vary_Psi} \\
    \frac{\delta}{\delta\omega^I} \ : \quad &D_\omega\Sigma^I = 0 \ ; \label{eq:vary_omega} \\
    \frac{\delta}{\delta\phi_a^I} \ : \quad &F^a\wedge\tilde\Sigma^I = \phi^a_J\,\tilde\Sigma^{(J}\wedge\Sigma^{I)} \ ; \label{eq:vary_psi} \\
    \frac{\delta}{\delta A^a} \ : \quad &D_A(\phi^a_I\tilde\Sigma^I) = 0 \ , \label{eq:vary_A}
\end{align}
where $D_\omega$/$D_A$ is the exterior covariant derivative w.r.t. the corresponding connection. The undetermined component of \eqref{eq:vary_Psi} is due to the fact that, according to \eqref{eq:Psi_symmetries}, we do not vary the trace of $\Psi^{IJ}$.

The constraint \eqref{eq:vary_Psi} on $\Sigma^I$ is the door through which metric geometry enters the present formalism. It has four branches of solutions \cite{DePietri:1998hnx,Freidel:1999rr} \emph{constructed from a vielbein} $e^A$, via the wedge product $(e\wedge e)^I \equiv e^{A_1}\wedge e^{A_2}$:
\begin{align}
    \Sigma^I = \pm (e\wedge e)^I \quad \text{or} \quad \tilde\Sigma^I = \pm (e\wedge e)^I \ . \label{eq:Sigma_e}
\end{align}
Regardless of the branch, eq. \eqref{eq:vary_omega} now becomes a torsion-free condition, which establishes $\omega^I$ as the spin-connection compatible with the vielbein $e^A$, and $R^I$ as its Riemann tensor. Now, let us tentatively take $\Sigma^I = (e\wedge e)^I$ as the ``correct'' branch of \eqref{eq:Sigma_e}, and plug it into the remaining equations \eqref{eq:vary_psi}-\eqref{eq:vary_A}. Eq. \eqref{eq:vary_psi} becomes:
\begin{align}
    F^a = \phi^a_I (e\wedge e)^I \ , \label{eq:F_psi}
\end{align}
which identifies the 0-forms $\phi^a_I$ as the components of the 2-form $F^a$. Eq. \eqref{eq:vary_A} becomes:
\begin{align}
    D_A\left(\tilde\phi^a_I (e\wedge e)^I\right) = 0 \ ,
\end{align}
which we now recognize via \eqref{eq:F_psi} as the YM field equation $D(\star F) = 0$. The only thing we're apparently missing is the Einstein equation. But this is actually contained in our solution $\Sigma^I = (e\wedge e)^I$ to eq. \eqref{eq:vary_Psi}! Recalling our definition $\tilde\Sigma_I = -M^{-1}_{IJ}V^J$, plugging in eqs. \eqref{eq:M}-\eqref{eq:V},\eqref{eq:F_psi} and solving for the Riemann curvature $R^I$, we have:
\begin{align}
    R^I = \left(\frac{\Lambda}{3}\,\eta^{IJ} + \Psi^{IJ} + \frac{\kappa}{2g^2}(\phi_a^I\phi_a^J + \tilde\phi_a^I\tilde\phi_a^J) \right)(e\wedge e)_J \ . \label{eq:Einstein}
\end{align}
Here, the 1st term establishes $4\Lambda$ as the Ricci scalar; the 2nd term establishes $\Psi^{IJ}$ as the Weyl curvature, while its non-Weyl-like extra components (the Ricci-like components of $\tilde\Psi^{IJ}$) are forced to vanish by the index symmetries of the Riemann tensor $R^I$; and, finally, the 3rd term establishes the traceless part of the Ricci tensor as $\kappa$ times the YM stress-energy tensor. Thus, we have precisely the Einstein equation, along with an identification of $\Psi^{IJ}$ as the Weyl curvature.

Now, let's return to consider the other branches of solutions \eqref{eq:Sigma_e} to eq. \eqref{eq:vary_Psi}. These amount to minus signs and/or Hodge duals on the factor of $(e\wedge e)^I$ in eqs. \eqref{eq:F_psi}-\eqref{eq:Einstein}. If we pick one of the ``Hodge-dual'' branches $\tilde\Sigma^I = \pm (e\wedge e)^I$, then eq. \eqref{eq:Einstein} doesn't define a valid Riemann tensor. In particular, the $\Lambda$ term becomes a ``Hodge dual of a Ricci scalar'', which is forbidden by the Riemann's symmetries, and isn't canceled by any other term in \eqref{eq:Einstein}. Therefore, these branches don't yield solutions of the entire system. The remaining branch $\Sigma^I = -(e\wedge e)^I$ yields a valid Riemann tensor in terms of its index symmetries, but with the YM stress-energy contributing to \eqref{eq:Einstein} with the wrong  sign (in fact, at the Lagrangian level, it's the kinetic energy of the graviton that has the wrong sign in this branch \cite{Mitsou:2019nlt}). Unfortunately, we are stuck with this branch as a valid solution to the equations of motion. However, it's cleanly separated from the ``correct'' branch $\Sigma^I = (e\wedge e)^I$, because the $\Lambda$ term in \eqref{eq:Einstein} produces a nonzero constant value for the Ricci scalar, with opposite signs for the two branches. Overall, we see that a nonzero $\Lambda$ plays three useful roles in this formalism:
\begin{enumerate}
    \item It ensures that the matrix \eqref{eq:M} is invertible near the ``vacuum'' $\phi^{Ia} = \Psi^{IJ} = 0$.
    \item It rules out the unphysical ``Hodge-dual'' branches $\tilde\Sigma^I = \pm (e\wedge e)^I$.
    \item It ensures that the physical branch $\Sigma^I = +(e\wedge e)^I$ is discontinuous with the ghost-like branch $\Sigma^I = -(e\wedge e)^I$.
\end{enumerate} 

\section{Complex, chiral Lagrangian}

Let us now present the chiral formulation of the same GR+YM theory. We separate the 6d space of bivectors (with indices $I,J,\dots$) into a pair of chiral, complex 3d subspaces (with indices $i,j,\dots$ and $i',j',\dots$). These are the eigenspaces of the Hodge dual, with eigenvalues $\pm i$ respectively. For any bivector $T^I$, the metrics $\eta_{IJ}$ and $\tilde\eta_{IJ}$ decompose as:
\begin{align}
  \begin{split}
    \eta_{IJ}T^I T^J &= \delta_{ij}T^i T^j + \delta_{i'j'}T^{i'}T^{j'} \ ; \\ 
    \tilde\eta_{IJ}T^I T^J &= i(\delta_{ij}T^i T^j - \delta_{i'j'}T^{i'}T^{j'}) \ ,
  \end{split}
\end{align}
where $\delta_{ij}$/$\delta_{i'j'}$ is the standard Euclidean metric. The Lorentz group in this language takes the form $SO(3,\bbC)$, with structure constants $\sqrt{2}\,\epsilon_{ijk}$ (and their complex conjugates $\sqrt{2}\,\epsilon_{i'j'k'}$).

Now, to construct a chiral formulation, we reduce our fundamental fields $(\omega^I,\Psi^{IJ},A^a,\phi^{Ia})$ to only the components along \emph{one} of the chiral subspaces. The other components are \emph{not} assumed to vanish -- they're simply no longer considered as part of the fundamental variables. Thus, the YM connection $A^a$ and its curvature \eqref{eq:F} stay the same; instead of $\omega^I$ and $\phi^{Ia}$, we consider only $\omega^i$ and $\phi^{ia}$ (which can be seen as a rearrangement of 6 real components into 3 complex ones); instead of the full GR curvature \eqref{eq:R}, we consider:
\begin{align}
     R^i \equiv d\omega^i + \frac{1}{\sqrt{2}}\,\epsilon^i{}_{jk}\,\omega^j\wedge\omega^k \ ;
\end{align}
finally, in place of $\Psi^{IJ}$, we take the symmetric traceless matrix $\Psi^{ij}$:
\begin{align}
   \Psi^{ij} = \Psi^{ji} \ ; \quad \Psi^i_i = 0 \ . \label{eq:Psi_symmetries_chiral}
\end{align}
Note that, unlike the case of $\omega^i$ and $\phi^{ia}$, this $\Psi^{ij}$ \emph{cannot} be viewed as a mere complex rearrangement of $\Psi^{IJ}$: it genuinely has fewer components, since we removed the mixed components $\Psi^{ii'}$ from consideration. Specifically, $\Psi^{ij}$ has 5 complex components, which (together with their complex conjugates $\Psi^{i'j'}$) form precisely the 10 real components of the Weyl curvature.

The chiral version of the Lagrangian \eqref{eq:L} reads:
\begin{align}
    \calL[\omega^i,\Psi^{ij},A^a,\phi^{ia}] = -M^{-1}_{ij}\,V^i\wedge V^j \ , \label{eq:L_chiral}
\end{align}
where:
\begin{align}
  M^{ij}[\Psi^{ij},\phi^{ia}] &= \frac{i}{\kappa}\left(\frac{\Lambda}{3}\,\delta^{ij} + \Psi^{ij}\right) - \frac{i}{g^2}\,\phi_a^i \phi_a^j \ ; \label{eq:M_chiral} \\ 
  V^i[\omega^i,A^a,\phi^{ai}] &= \frac{1}{\kappa}R^i - \frac{1}{g^2}\,\phi_a^i F^a \ . \label{eq:V_chiral}   
\end{align}
Denoting $M^{-1}_{ij}V^j \equiv -i\Sigma_i$ (so that $\Sigma_i$ will be the chiral components of $\Sigma_I$ from the previous section), the equations of motion read:
\begin{align}
    \frac{\delta}{\delta\Psi^{ij}} \ : \quad &\Sigma^i\wedge\Sigma^j \sim \delta^{ij} \ ; \label{eq:vary_Psi_chiral} \\
    \frac{\delta}{\delta\omega^i} \ : \quad &D_\omega\Sigma^i = 0 \ ; \label{eq:vary_omega_chiral} \\
    \frac{\delta}{\delta\phi_a^i} \ : \quad &F^a\wedge\Sigma^i = \phi^a_j\,\Sigma^j \wedge\Sigma^i \ ; \label{eq:vary_psi_chiral} \\
    \frac{\delta}{\delta A^a} \ : \quad &D_A(\phi^a_i\Sigma^i) = 0 \ , \label{eq:vary_A_chiral}
\end{align}
The simplicity constraint \eqref{eq:vary_Psi_chiral} now has only one branch of solutions $\Sigma^i \sim (e\wedge e)^i$, in terms of a \emph{complex} vielbein (or, equivalently, a complex Urbantke metric \cite{Urbantke:1984eb}). The four real branches \eqref{eq:Sigma_e} can be expressed as $\Sigma^i = c(e\wedge e)^i$, where the vielbein is now real, and the numerical factor $c$ is $\pm 1$ or $\pm i$. The ``correct'' real branch corresponds to $c=1$. The complete basis of 2-forms $\Sigma^I$ is now spanned by $\Sigma^i$ together with their complex conjugates $\Sigma^{i'}$.

With this understanding, eq. \eqref{eq:vary_omega_chiral} is again the torsion-free condition that fixes $\omega^i$ to be compatible with the vielbein. Eq. \eqref{eq:vary_psi_chiral} identifies the 0-forms $\phi^i_a$ as the chiral components of the 2-form $F^a$, as in:
\begin{align}
    F^a = \phi^a_i\Sigma^i + \phi^a_{i'}\Sigma^{i'} \ . \label{eq:F_psi_chiral}
\end{align}
Eq. \eqref{eq:vary_A_chiral} then implies the YM field equation $D(\star F) = 0$, in linear combination with the Bianchi idendity $DF = 0$. Finally, unpacking the relation $M^{-1}_{ij}V^j = -i\Sigma_i$ and solving for $R^i$, we obtain the chiral analogue of \eqref{eq:Einstein}:
\begin{align}
    R^i = \left(\frac{\Lambda}{3}\,\delta^{ij} + \Psi^{ij} \right)\Sigma_j + \frac{\kappa}{g^2}\,\phi_a^i\phi_a^{i'}\,\Sigma_{i'} \ , \label{eq:Einstein_chiral}
\end{align}
which is the correct Einstein equation analogously to \eqref{eq:Einstein}, with $\Psi^{ij}$ identified as (the chiral half of) the Weyl curvature.

\subsection{Self-dual sector}

One virtue of chiral Lagrangians such as \eqref{eq:L_chiral}-\eqref{eq:V_chiral} is that they contain the theory's self-dual sector as a limit \cite{Krasnov:2016emc}. To arrive at the self-dual sector, we must complexify the fields, so that $\omega^{i'},\Psi^{i'j'},\phi^{i'a}$ become independent from $\omega^i,\Psi^{ij},\phi^{ia}$, rather than their complex conjugates. We then take the limit in which the field strengths of one chirality, i.e. $\Psi^{ij}$ and $\phi^{ia}$, are small. In other words, we expand the Lagrangian \eqref{eq:L_chiral}-\eqref{eq:V_chiral} in powers of $\Psi^{ij}$ and $\phi^{ia}$. The zeroth-order term $\calL_0 = \frac{3i}{\kappa\Lambda}R_i\wedge R^i$ is topological, and can be discarded. The self-dual sector is then contained in the first-order terms:
\begin{align}
   \begin{split}
     &\calL_{\text{self-dual}}[\omega^i,\Psi^{ij},A^a,\phi^{ia}] \\
     &\quad = -\frac{9i}{\kappa\Lambda^2}\,\Psi_{ij}R^i\wedge R^j - \frac{6i}{g^2\Lambda}\,\phi_{ia}R^i\wedge F^a \ .
   \end{split} \label{eq:L_SD}
\end{align}
This Lagrangian describes self-dual gravitational and YM fields, encoded in $\omega^i$ and $A^a$, along with linearized anti-self-dual perturbations, encoded in $\Psi^{ij}$ and $\phi^{ia}$. In this limit, the anti-self-dual perturbations don't backreact on the nonlinear self-dual fields. In particular, this implies that the YM fields don't backreact on the geometry, since the YM stress-energy tensor contains a factor of $\phi^{ia}$ -- see \eqref{eq:Einstein_chiral}.

\section{Derivation from a Plebanski-type Lagrangian}

Having analyzed the new Lagrangians \eqref{eq:L},\eqref{eq:L_chiral} on their own terms, let us now demystify them by ``integrating in'' the 2-forms $\Sigma$ as independent variables. In the non-chiral case, this gives the Lagrangian:
\begin{align}
   &\calL[\Sigma^I,\omega^I,\Psi^{IJ},A^a,\phi_a^I] = \frac{1}{2}M_{IJ}\tilde\Sigma^I\!\wedge\tilde\Sigma^J + V_I\wedge\tilde\Sigma^I \\
   \begin{split}
     &= \frac{1}{\kappa}\,\tilde\Sigma^I \wedge\left(R_I - \frac{\Lambda}{6}\,\Sigma_I - \frac{1}{2}\Psi_{JI}\Sigma^J \right) \\
     &\qquad - \frac{1}{g^2}\,\phi^a_I\,\tilde\Sigma^I \wedge\left(F^a - \frac{1}{2}\phi_J^a\,\Sigma^J \right) \ .
   \end{split} \label{eq:L_unpacked}
\end{align}
Varying this w.r.t. $\Sigma^I$, we obtain $\tilde\Sigma_I = -M_{IJ}^{-1}V^J$ as a field equation. Plugging this back into the Lagrangian to eliminate $\Sigma^I$, we recover \eqref{eq:L}. The chiral Lagrangian \eqref{eq:L_chiral} arises in the same way from:
\begin{align}
    &\calL[\Sigma^i,\omega^i,\Psi^{ij},A^a,\phi_a^i] = -M_{ij}\Sigma^i\wedge\Sigma^j + 2iV_i\wedge\Sigma^j \\
    \begin{split}
        &= \frac{2i}{\kappa}\,\Sigma^i\wedge \left(R_i - \frac{\Lambda}{6}\,\Sigma_i - \frac{1}{2}\Psi_{ij}\Sigma^j \right) \\
        &\qquad - \frac{2i}{g^2}\,\phi_i^a\Sigma^i\wedge \left( F_a - \frac{1}{2}\phi^a_j\Sigma^j \right) \ .
    \end{split} \label{eq:L_unpacked_chiral}
\end{align}
Unlike our main Lagrangians \eqref{eq:L},\eqref{eq:L_chiral}, the unpacked versions \eqref{eq:L_unpacked},\eqref{eq:L_unpacked_chiral} take the standard form ``GR term + YM term''. In particular, the GR term in \eqref{eq:L_unpacked_chiral} is just the Plebanski Lagrangian, while the YM term is the Chalmers-Siegel Lagrangian (adapted to curved spacetime, as in \cite{Capovilla:1991qb}). The corresponding terms in \eqref{eq:L_unpacked} are just the non-chiral versions of these.

\section{Outlook}

In this paper, we presented and analyzed two Lagrangians \eqref{eq:L}-\eqref{eq:V} and \eqref{eq:L_chiral}-\eqref{eq:V_chiral} that describe a GR+YM theory, with the same kind of variables in both sectors, playing maximally similar roles. The choice of fundamental variables -- 1-forms and 0-forms -- was inspired by higher-spin theory. There, a Lagrangian formulation similar to \eqref{eq:L_SD} has recently been given for higher-spin generalizations of self-dual GR and self-dual YM \cite{Krasnov:2021nsq}. The present paper's results, which cover full GR+YM theory, may contain useful clues for a better formulation of full higher-spin theory. 

Back in GR+YM, it would be interesting to see whether the double-copy structure of our Lagrangians leads to some interesting consequences in scattering problems, such as the usual S-matrix in the $\Lambda\to 0$ limit \cite{Delfino:2012aj}, or de Sitter static-patch amplitudes directly at $\Lambda\neq 0$ \cite{Albrychiewicz:2021ndv,Neiman:2023bkq}. In this context, we expect the chiral formulation \eqref{eq:L_chiral}-\eqref{eq:V_chiral} to be superior. Indeed, in perturbative calculations, there is no harm in using complex fields or actions. Moreover, as remarked above, the chiral version allows us to view the full theory as a perturbative expansion over its self-dual sector \eqref{eq:L_unpacked_chiral}; in scattering problems, this corresponds to organizing all the amplitudes as perturbations around the MHV sector.

Finally, it would be wonderful if the present formalism can be extended to include spin-$\frac{1}{2}$ matter, perhaps by way of supersymmetry. In the context of the Plebanski-type Lagrangian \eqref{eq:L_unpacked_chiral}, some steps in this direction were made in \cite{Capovilla:1991qb,Capovilla:1991kx}, but it appears that new ideas are still needed.

\begin{acknowledgments}
I am grateful to Keith Glennon, Julian Lang and Kirill Krasnov for discussions. I am supported by the Quantum Gravity Unit of the Okinawa Institute of Science and Technology Graduate University (OIST).
\end{acknowledgments}

\end{document}